\documentclass[conference]{IEEEtran}

\usepackage{cite}
\usepackage{amsmath,amssymb,amsfonts}
\usepackage{algorithmic}
\usepackage{graphicx}
\usepackage{textcomp}
\usepackage{xcolor}
\usepackage{hyperref}
\usepackage{soul}
\def\BibTeX{{\rm B\kern-.05em{\sc i\kern-.025em b}\kern-.08em
    T\kern-.1667em\lower.7ex\hbox{E}\kern-.125emX}}
\begin{document}

\title{Earthquake Simulation Project\\
\thanks{*Equal Contribution.}
}

\author{\IEEEauthorblockN{Palak Chawla}
\IEEEauthorblockA{\textit{Computer Engineering} \\
\textit{University of Central Florida}\\
Orlandlo, Florida, USA \\
pa973175@ucf.edu}
}

\maketitle

\begin{abstract}
This paper presents a seismic activity simulator that models the effects of fault lines on surface pressure. This project uses C programming to create a fully interactive learning resource intended to educate users on the mechanics of earthquakes. The motivation behind this project is to make studying seismic activity more accessible, engaging and cost effective.  
\end{abstract}


\section{Introduction}
\label{sec:motivation}
Earthquakes are highly destructive natural disasters that cause over tens of thousands of deaths each year\cite{ourworldindata_earthquake_deaths}, as well as significant property damage, and tens of billions in dollars of financial loss in the United States alone\cite{usgs_fema_earthquake_risk}. It is imperative to understand the workings of earthquakes in order to further preventative measures to minimize damages. This simulator was created in order to facilitate this learning and lead to breakthroughs in earthquake forecasting and prevention technology. This simulator uses C programming to allow users to input geographical area and fault lines, which are used to provide a visual, complete with color coding to help the user identify patterns within the output and understand seismology. This paper covers the development of the code, including the explanations of the functions and algorithms used, explanations of edge cases, and related projects in earthquake forecasting technology.

\section{Related Works}
\label{sec:literature}
The Convolutional Recurrent Model for Earthquake Identification and Magnitude Estimation was created at the Frankfurt Institute for Advanced Studies and Goethe University in Frankfurt, by Megha Chakraborty, Darius Fenner, Wei Li, Johannes Faber, Kai Zhou, Georg Rümpker, Horst Stöcker, and Nishtha Srivastava. \cite{https://doi.org/10.1029/2022JB024595} This project aimed to enhance the accuracy of earthquake identification and magnitude estimation. They accomplished this by using large datasets of seismic activity and using machine learning techniques to teach the model to recognize patterns and predict earthquake occurrences. The Convolutional Recurrent Model for Earthquake Identification and Magnitude Estimation is an effective tool to predict earthquakes since it is based on real life data, whereas the project in this paper is more suitable to be used as a learning resource to allow one to understand seismic patterns.\\

The Uniform California Earthquake Rupture Forecast was a project done at Columbia University under Dr. Bruce E. Shaw.\cite{ucerf_appendixE}This project aimed to enhance the accuracy and physical realism of seismic hazard modeling through large-scale, physics-based earthquake simulators. They accomplished this goal using advanced simulation techniques with physically informed modeling to produce more accurate reliable tools for earthquake hazard analysis. This project is more similar to the one in this paper as it is also a physics-based simulator seeking to improve the understanding of fault line and stress level interactions.

\section{Methodology}
\vspace{1em}

\subsection{Overview}

The program is organized into 8 functions. Print fault map, print surface map, vertical, horizontal, circular, point to point, update surface, and main. The print fault map function prints the fault map array, the print surface map prints the surface map array, the vertical function adds a user defined vertical line on to the fault map array, the horizontal function adds a user defined horizontal line onto the fault map array, the circular line adds circle onto the fault map array based on a user defined center point, the point to point function adds a line to the fault line function based on two user defined points, the update surface function when called in main after the fault map has been fully updated adds random stress to the surface stress array based on whether or not the point is on a fault. The code relies on 2D arrays to represent the geographical area to draw the fault lines on, as well as to simulate the different stress levels. There were two main algorithms that were used to compute the shapes to draw on the fault map. The Bresenham’s line algorithm, and the midpoint circle algorithm. The main function starts by providing the user with a menu to decide what shape they would like to add to the fault grid. This code utilizes the standard input output library to facilitate the printing of output onto the screen as well as taking in the input needed to run the simulation. Additionally, the standard library is utilized to use the random function to simulate surface stress values, as well as the time library to seed time within the random number generator. Lastly, the universal standard library is used to employ the sleep function to slow down the output, making it clearer for the user to see how the simulation updates. 

\subsection{Functions}

\subsubsection{Print Fault Map}
The print fault map function prints out the 2D grid on which the fault lines are entered. It prints fault line points in red using the number 1, and the non fault lines using the number 0 in white. As the user enters the different shapes which are stored on the array, this function prints. 

\vspace{1em}

\subsubsection{Print Surface Map}
The print surface map function prints out the 2D grid on which the surface pressure is updating. It colors the low stress points in green, medium stress points in yellow, high stress points in red, and earthquake points in blue. This is done by printing these points from the surface stress array in colors using the ANSI escape codes that are defined as symbolic constants on top of the code. 

\vspace{1em}

\subsubsection{Vertical}
The vertical function takes in x coordinate input, and starts by checking if the input is valid. If the point is valid, a vertical line is drawn at that x coordinate on the fault map array. If the input is not within range of the grid, an error message is printed. 

\vspace{1em}

\subsubsection{Horizontal}
The horizontal function takes in y coordinate input, and starts by checking if the input is valid. If the point is valid, a horizontal line is drawn at that y coordinate on the fault map array. If the input is not within range of the grid, an error message is printed. 

\vspace{1em}

\subsubsection{Circular}
The circular function takes in input x and y of a coordinate, which is the center point of the circle. The function starts by ensuring that the value is within the range of the grid, printing an error message if not, then executes the midpoint circle algorithm to plot a circle on the graph. This requires starting at the highest point of the circle which is at the x coordinate 0 and the y coordinate is the radius of the circle. Then moving horizontally or diagonally based on calculating the error. The circle is divided into 8 equal sections, called octants.  While we draw the circle, the drawing gets mirrored over the full circle, because of the if statements implemented to do so for each octant. This way, an accurate circle can be drawn using minimal calculations. A check is then done to see if half the circle is complete, implying the rest is done too due to the mirroring. Once the circle is complete, the fault map is updated. 

\vspace{1em}

\subsubsection{Point to Point}
The point to point function uses the Bresenham's line algorithm to efficiently draw a line between two user-defined coordinates. The user is asked for the x and y for both these coordinates, which are used by the algorithm to plot the next point either horizontally, vertically, or diagonally based on error moving forward. We move forward from both the points to eventually stop when the points meet in the middle forming the full line. Once this line is complete, the fault map is updated. 

\vspace{1em}

\subsubsection{Update Surface}
The update surface function assigns a random value to each point on the surface stress array based on the fault line array. A stress between -5 and 5 is added to non-fault lines points, with a condition to set any negative stress to 0, since we do not want the stress to go below that. The fault points get a stress of 0-10 added to them as the simulation updates. This function includes a command to clear the screen each time an updated surface stress map is printed, giving the illusion of a real time refresh. Between each updated surface map, the sleep function is used to slow down the updates by 1 second. Since this simulation is meant as a learning tool, this is a vital feature to allow the user to better observe how the simulation changes with each update, and what kinds of patterns emerge. Finally, this function prints a message indicating an earthquake once one occurs. 

\vspace{1em}

\subsubsection{Main}
The main function is used to call the rest of the functions and direct the flow of the overall simulation. Firstly the fault line shapes menu is printed within the main, and it takes in user input on what fault line shapes to draw, calling the necessary function when needed. The print fault map and print surface map function are both called within main after the user decided to start the simulation. Then the main function runs a loop to call the update surface map function until the desired amount of earthquakes, defined as a symbolic constant, have been reached. Finally, the main function returns 0 to indicate the successful excecution of the program. 

\section{Edge Cases}
Within this earthquake simulator, there are many edge cases that have been accounted for to improve the accuracy of this program. Each shape function comes with a check to ensure that input is within the defined range and prints an error message if the user tries to draw a fault line outside the defined geographical area. Another edge case is with the point to point line. The program is successfully able to draw horizontal, vertical and steep lines using the point to point line function due to the precision of Bresenham's line algorithm. Furthermore, an edge case relating to circles that arises is printing a cropped circle when the center point coordinates are within bounds, but the circle does not fit within the defined range.  Using the midpoint circle algorithm, this program is correctly able to draw a cropped circle from a center point within the grid even if part of it is cut off, without printing off the grid or wrapping around to the other side. Handling these edge cases is imperative to allows my simulator to be a successful educational resource, offering users accurate results. 

\section{Conclusion}
This paper presented a C-based earthquake simulation tool that models stress accumulation across fault lines within a customizable 2D grid. The simulator visualizes fault shapes and stress levels in real time using ANSI escape codes, providing an intuitive and educational interface for understanding seismic behavior. The implementation includes functionality for various fault shapes, stress distribution, and earthquake triggering mechanisms. By addressing input constraints and edge cases, the program ensures consistent performance across diverse scenarios allowing for a reliable and accessible educational tool. Future work may include incorporating real-world geological data, similar to the UCERF project to make this simulator more realistic, expanding the uses of this project from an educational tool to one that could further seismic research. 


\begin{figure*}[!t]
  \centering
  \includegraphics[width=\textwidth]{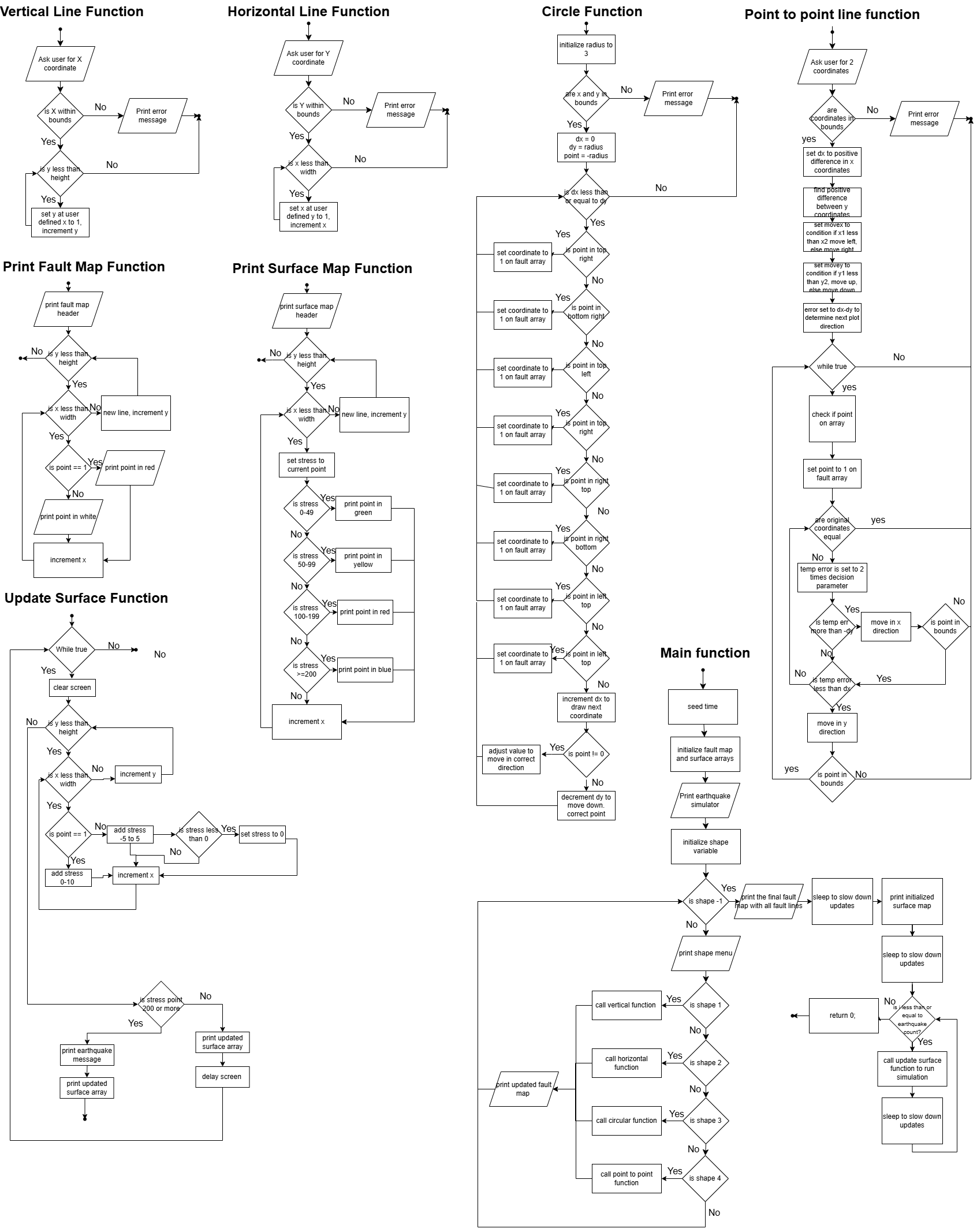}
\end{figure*}

\small
\bibliographystyle{IEEEtran}
\bibliography{main}

\end{document}